\documentclass[12pt]{article}

\begin{document}


\newcommand{\bb}{\begin{equation}}
\newcommand{\ee}{\end{equation}}
\newcommand{\bbb}{\begin{eqnarray}}
\newcommand{\eee}{\end{eqnarray}}
\newcommand{\diag}{\mbox{diag }}
\newcommand{\Str}{\mbox{STr }}
\newcommand{\Tr}{\mbox{Tr }}
\newcommand{\Det}{\mbox{Det }}
\newcommand{\C}[2]{{\lk [{#1},{#2}\re ]}}
\newcommand{\AC}[2]{{\lk \{{#1},{#2}\re \}}}
\newcommand{\kk}{\hspace{.5em}}
\newcommand{\vc}[1]{\mbox{$\vec{{\bf #1}}$}}
\newcommand{\mc}[1]{\mathcal{#1}}
\newcommand{\del}{\partial}
\newcommand{\lk}{\left}
\newcommand{\ave}[1]{\mbox{$\langle{#1}\rangle$}}
\newcommand{\re}{\right}
\newcommand{\pd}[1]{\frac{\del}{\del #1}}
\newcommand{\pdd}[2]{\frac{\del^2}{\del #1 \del #2}}
\newcommand{\Dd}[1]{\frac{d}{d #1}}
\newcommand{\sech}{\mbox{sech}}
\newcommand{\pref}[1]{(\ref{#1})}

\newcommand
{\sect}[1]{\vspace{20pt}{\LARGE}\noindent
{\bf #1:}}
\newcommand
{\subsect}[1]{\vspace{20pt}\hspace*{10pt}{\Large{$\bullet$}}\mbox{ }
{\bf #1}}
\newcommand
{\subsubsect}[1]{\hspace*{20pt}{\large{$\bullet$}}\mbox{ }
{\bf #1}}

\def\ie{{\it i.e.}}
\def\eg{{\it e.g.}}
\def\cf{{\it c.f.}}
\def\etal{{\it et.al.}}
\def\etc{{\it etc.}}

\def\e{{\mbox{{\bf e}}}}
\def\AA{{\cal A}}
\def\BB{{\cal B}}
\def\CC{{\cal C}}
\def\DD{{\cal D}}
\def\EE{{\cal E}}
\def\FF{{\cal F}}
\def\GG{{\cal G}}
\def\HH{{\cal H}}
\def\II{{\cal I}}
\def\JJ{{\cal J}}
\def\KK{{\cal K}}
\def\LL{{\cal L}}
\def\MM{{\cal M}}
\def\NN{{\cal N}}
\def\OO{{\cal O}}
\def\PP{{\cal P}}
\def\QQ{{\cal Q}}
\def\RR{{\cal R}}
\def\SS{{\cal S}}
\def\TT{{\cal T}}
\def\UU{{\cal U}}
\def\VV{{\cal V}}
\def\WW{{\cal W}}
\def\XX{{\cal X}}
\def\YY{{\cal Y}}
\def\ZZ{{\cal Z}}

\def\sinh{{\rm sinh}}
\def\cosh{{\rm cosh}}
\def\tanh{{\rm tanh}}
\def\sgn{{\rm sgn}}
\def\det{{\rm det}}
\def\trace{{\rm Tr}}
\def\exp{{\rm exp}}
\def\sh{{\rm sh}}
\def\ch{{\rm ch}}

\def\ell{{\it l}}
\def\str{{\it str}}
\def\lp{\ell_{{\rm pl}}}
\def\blp{\overline{\ell}_{{\rm pl}}}
\def\ls{\ell_{{\str}}}
\def\bls{{\bar\ell}_{{\str}}}
\def\bM{{\overline{\rm M}}}
\def\gs{g_\str}
\def\gym{{g_{Y}}}
\def\geff{g_{\rm eff}}
\def\eff{{\rm eff}}
\def\r11{R_{11}}
\def\kel{{2\kappa_{11}^2}}
\def\kten{{2\kappa_{10}^2}}
\def\lpten{{\lp^{(10)}}}
\def\alp{{\alpha '}}
\def\alpe{{{\alpha_e}}}
\def\le{{{l}_e}}
\def\aleff{{\alp_{eff}}}
\def\sqaleff{{\alp_{eff}^2}}
\def\tgs{{\tilde{g}_s}}
\def\talp{{{\tilde{\alpha}}'}}
\def\tlp{{\tilde{\ell}_{{\rm pl}}}}
\def\tr11{{\tilde{R}_{11}}}
\def\wtilde{\widetilde}
\def\what{\widehat}
\def\hlp{{\hat{\ell}_{{\rm pl}}}}
\def\hr11{{\hat{R}_{11}}}
\def\hf{{\textstyle\frac12}}
\def\coeff#1#2{{\textstyle{#1\over#2}}}
\def\CY{Calabi-Yau}
\def\lessapprox{\;{\buildrel{<}\over{\scriptstyle\sim}}\;}
\def\greaterapprox{\;{\buildrel{>}\over{\scriptstyle\sim}}\;}
\def\inbar{\,\vrule height1.5ex width.4pt depth0pt}
\def\IC{\relax\hbox{$\inbar\kern-.3em{\rm C}$}}
\def\IR{\relax{\rm I\kern-.18em R}}
\def\IP{\relax{\rm I\kern-.18em P}}
\def\Z{{\bf Z}}
\def\R{{\bf R}}
\def\One{{1\hskip -3pt {\rm l}}}
\def\sst{\scriptscriptstyle}
\def\osc{{\rm\sst osc}}
\def\lam{\lambda}
\def\lc{{\sst LC}}
\def\pr{{\sst \rm pr}}
\def\cl{{\sst \rm cl}}
\def\D{{\sst D}}
\def\bh{{\sst BH}}
\def\vev#1{\langle#1\rangle}

\newcommand{\Sym}{\mbox{{\bf Sym}}}
\newcommand{\Tless}{\mbox{{\bf Traceless}}}

\input{epsf}

\begin{titlepage}
\rightline{CLNS 01/1725}

\rightline{hep-th/0102200}

\vskip 2cm
\begin{center}
\Large{{\bf 
On D0 brane polarization \\ by tidal forces
}}
\end{center}

\vskip 2cm
\begin{center}
Vatche Sahakian\footnote{\texttt{vvs@mail.lns.cornell.edu}}
\end{center}
\vskip 12pt
\centerline{\sl Laboratory of Nuclear Studies}
\centerline{\sl Cornell University}
\centerline{\sl Ithaca, NY 14853, USA}

\vskip 2cm

\begin{abstract}
Gravitational tidal forces may induce polarization of D0 branes, in analogy
to the same effects arising in the context of constant background gauge fields.
Such phenomena can teach us about the correspondence between smooth curved spacetime
and its underlying non-commutative structure. However, unlike 
polarization by gauge fields, the gravitational counterpart involves
concerns regarding the classical stability of the corresponding polarized states.
In this work, we study this issue with respect to the solutions presented in
hep-th/0010237 and find that they are classically unstable.
The instability however appears with intricate features with all but a few 
decay channels being lifted. Through a detailed analysis, we then argue that
these polarized states may be expected to be long-lived in a regime where
the string coupling is small and the number of D0 branes is large.
\end{abstract}

\end{titlepage}
\newpage
\setcounter{page}{1}

\section{Introduction}
\label{intro}

Non-perturbative dynamics in string theory leads one to believe that
spacetime, when probed near the Planck scale,
is to acquire exotic structure; one that appears to involve
non-commutativity of 
coordinates and an associated foam-like picture for space. The focus of
most recent programs has been to understand
phenomena of non-commutativity of spacetime coordinates 
as induced by certain supergravity gauge fields in the background of flat 
space~\cite{SWNC}-\cite{CORSCHI}. 
These situations present
excellent laboratories to test one's intuition with respect to
non-commutative geometry, decoupled from gravitational complications.
We still however lack a fundamental understanding 
correlating notions of smoothly curved spacetime, as it arises
in the context of familiar gravitational dynamics, and the underlying
Planck scale fuzziness (for recent work addressing this question, 
see~\cite{TAYLORMAT,CORSCHI}).

In a previous work~\cite{VVSD}, 
we attempted to address this issue by trying to understand
how information about curved spaces, organized for example
in derivatives of the metric at a given point, 
can get encoded into D brane coordinates; given that D branes are known to be natural
probes of short length scales. 
We focused on D0 branes immersed in curved spaces, and used
the results 
of~\cite{WATIDBI,TAYLORMAT,MEYERS,GARMEY1,GARMEY2,MUKHISURYAN,TATAR,MINAS} 
to explore the resulting dynamics
through the non-abelian Dirac-Born-Infeld (DBI) action.
We identified extrema
of this action corresponding to polarized D0 branes; 
polarization being induced by gravitational tidal forces. 
From this setting, we learned about 
the balance of forces between the smooth gravitational background fields
and the inter-brane interactions described through non-commuting matrices. And
we found a map between data about the smooth background geometry and
the structure constants of algebras that characterize 
the D0 brane matrix coordinates. 

An issue of concern in this picture is the classical stability of
the polarized configurations. In the scenario of interest,
the system is described by a bottomless 
potential, with directions corresponding to diagonal matrices that 
can lower the energy indefinitely. Physically, these directions correspond
to channels for 
evaporation into D0 branes that do not have any collective coherence.
The question is then whether the extrema of~\cite{VVSD} are
local minima of the potential energy; hence, corresponding to metastable
configurations; or whether they are classically unstable states and would
decay promptly into D0 brane gas.

The purpose of the current work is to address this issue of stability.
We will find that the relevant extrema of the DBI action
are saddle points of the potential energy, and are hence classically unstable.
The instability is of a peculiar nature, with most directions
in phase space leading to stable modes; 
including diagonal matrix channels. Using this information,
we then estimate the time the system spends near these extrema and argue
that, for large number of D0 branes, and small enough string coupling,
this timescale can be very long; that is long enough to make the polarized states
physically relevant to the dynamics of the system. The basic idea
is that small string
coupling increases the inertia of the D0 branes; and that
the available phase space leading to decay, as a fraction of the total
phase space, decreases in volume as $1/N^2$ for one particular 
class of polarized states. 

In Section~\ref{prelim}, we briefly review the problem at hand. 
In Sections~\ref{modes} and~\ref{shape},
we present the spectrum of the oscillators resulting from perturbing the solutions
of concern. In Sections~\ref{time} and~\ref{reflections}, 
we analyze the dynamics of decay,
present the main conclusions of this work, and assess the relevance of
the discussion to understanding non-commutativity in curved spaces.
Finally, Section~\ref{details} contains some of the details of the
calculations.

\section{Preliminaries}
\label{prelim}

We consider 
a static field configuration of
type IIA supergravity
involving
non-trivial metric, dilaton and RR two-form field strength;
all other fields being assumed zero. A collection of $N$
D0 branes is to propagate in this space as a probe of the geometry.
We concentrate on a regime where back-reaction effects and
radiation from the accelerating D0 branes can be ignored. 
The dynamics may be described by the non-Abelian $0+1$ dimensional 
DBI action
with gauge group U(N). Expanding this action to quadratic order 
in $\alp$,
we can separate the center of mass 
motion of the D0 branes from the remaining dynamics in the SU(N). This U(1) sector then
describes the collective motion of the D0 branes as they fall 
along a geodesic trajectory 
in the given background geometry 
(see~\cite{VVSD} for the details).
Choosing Fermi normal coordinates~\cite{FERMI}, we focus on the SU(N) dynamics as seen
by an observer falling with the center of mass of the D0 branes. 
We look for solutions in SU(N) describing D0 branes polarized into D2
branes under the influence of tidal-like forces that pull apart the
configuration. The energy is given by
\bb\label{energy}
E=
\frac{\lk(2 \pi \ls^2 \re)^2}{\gs\ls} \Str \lk\{ 
\frac{1}{2} \dot{\Phi}^n \dot{\Phi}^n
+M_{nm} \Phi^n \Phi^m
-\frac{1}{4} \C{\Phi^n}{\Phi^m} \C{\Phi^n}{\Phi^m}\re\}\ ,
\ee
where
\bb
M_{nm}\equiv -\frac{1}{4} G_{00,nm}-\frac{1}{2} C^{(1)}_{0,nm}\ .
\ee
$C^{(1)}$ is the one form RR potential, and the time-time component of
the metric $G_{00}$ is given with respect to the string frame metric as
\bb
G_{00}=e^{-2\phi} G_{00}^{str}\ .
\ee
The $\Phi^n$'s are time-dependent 
traceless Hermitian matrices, with $n=1,\ldots,9$.
The equation of motion is
\bb\label{eom}
{\ddot{\Phi}}^n+2 M_{nm} \Phi^m +\C{\Phi^m}{\C{\Phi^m}{\Phi^n}}=0\ .
\ee

We specialize to a background geometry of the form
\bb
M_{ij}=\delta_{ij} M\ ,\ \ \ 
M_{ab}=\delta_{ab} M'\ \ \ \mbox{with}\ \ \ M<0\ .
\ee
Here, $i,j=1,2,3$ and $a,b=4,\ldots,9$; the nine dimensional space 
is hence split into two subspaces.
For scenarios involving 
somewhat more general backgrounds, the reader is referred to~\cite{VVSD}.
With the present choice, we are focusing on configurations that can
potentially carry D2 brane charge; the D2 branes extending in the three
directions labeled by $i,j$.
And the isotropy of the background within each subspace is chosen to
avoid unnecessary clutter; the general conclusions we will reach are 
expected to be insensitive to anisotropy. The condition $M<0$ implies that
the tidal forces are pulling apart the D0 branes, as opposed to
exerting an inward pressure.

There are several classes of solutions to~\pref{eom}. A dynamical 
scenario is realized by choosing the $\Phi^n$ matrices in the
Cartan subalgebra of SU(N); \ie\ diagonal traceless matrices. With $M<0$,
this describes the D0 branes evaporating to infinity, without any 
collective coherence.
This is an uninteresting scenario that
appears to be the preferred channel 
for the evolution of the system.

A closer inspection of equation~\pref{eom} leads to other
solutions that describe polarized configurations; perhaps
metastable states that the system may prefer given 
the appropriate initial conditions.
One such realization describes a time-independent non-commutative sphere~\cite{VVSD}
\bb\label{fuzzy}
\Phi^i=\pm \frac{\sqrt{-M}}{2} \tau^i\ ,\ \ \ 
\Phi^a=0\ ,
\ee
which couples to the D2 brane three-form potential.
The $\tau^i$'s are the SU(2) Pauli matrices in an $n\times n$ representation
embedded in $N$ dimensional SU(N) matrices (with $n\leq N$); 
with our choice of normalization, the algebra is
\bb
\C{\tau^i}{\tau^j}=2\ i\ \varepsilon_{ijk} \tau^k\ .
\ee 
This D2 brane extends in the three dimensional isotropic 
subspace of the nine dimensional space, and has size~\cite{WATISPHERE,VVSD}
\bb\label{size}
\frac{r}{2 \pi \alp}=\lk( -\frac{3 M}{4}\re)^{1/2}\ \lk( n^2-1 \re)^{1/2}\ .
\ee
Consistency of our approach requires that this size be much smaller than
the length scale set by the background field $M$; so that, locally,
the geometry is almost flat. This statement interestingly can be written as
a bound on $n$
\bb\label{Nbound}
n\ll \frac{1}{M \alp}\equiv \AA \ ,
\ee
where $\AA$ can be thought of as a measure of area constructed from the 
local curvature scale in Planck units. We will refer to the $n=N$ case
as the maximal non-commutative sphere.

Furthermore, the situation we consider necessarily
involves a ``slow'' evolution of the center of mass of the D0 branes
in the background geometry. It is well known that the Fermi normal 
coordinates are a good approximation of the local structure of space 
when time derivatives of the metric are much smaller than space derivatives~\cite{FERMI}.
This may for example be achieved by a judicious choice of initial conditions.
Crudely speaking, we need to focus on
parts of the history of the configuration where the system moves at small 
``speeds'' $L/T$; where $L$ is the local length scale set by $M\sim 1/L^2$, 
and $T$ is the typical proper-time scale over which $M$ evolves. As the D0 branes
will see the fields about them vary slowly in this sense, static solutions
like~\pref{fuzzy} may be considered, and the adiabatic evolution of the 
non-commutative sphere 
is described (up to issues regarding its stability that we will address later)
by introducing the time dependence through
the background field $M\rightarrow M(t)$ in~\pref{fuzzy}. Naturally, when we
refer to a configuration as being long-lived or metastable, 
we imply that its lifetime is of order $T$, not $L$.
A system that lives for timescales of order $L$ is short-lived; it would consist of
a momentary nucleation in the history of the evolution of the center of mass.
The hierarchy between the scales $L$ and $T$ will play a crucial role
in our analysis of the stability of the solution~\pref{fuzzy}.

The energy of this spherical D2 brane is
\bb\label{v1}
V_{\mbox{sph}}=-\frac{\lk(2 \pi \ls^2 \re)^2}{8 \gs\ls} n\lk(n^2-1\re) M^2\ ;
\ee
while zero energy corresponds to the configuration where all the D0 branes
are sitting on top of each other at the origin of the coordinate system.
Hence, lowest energy in this set of solutions is achieved when $n=N$. 
We may also view the system from the point of view of the D2 brane
world-volume theory; the latter is characterized by a scale of
non-commutativity $\sqrt{\theta}\sim \ls^2/L$, and an IR cutoff
$\Sigma\sim r$.

It will be instructive to consider yet another class of static solutions
to~\pref{eom}. In particular, we can easily write down the alternative 
configurations
\bb\label{disc}
\Phi^1=0\ ,\ \ \ 
\Phi^2=\pm\sqrt{\frac{-M}{2}}\tau^2\ ,\ \ \ 
\Phi^3=\pm\sqrt{\frac{-M}{2}}\tau^3\ ,\ \ \ \Phi^a=0\ ;
\ee
with all permutations of the signs and of the
Pauli matrices being other possible realizations. We will call these
non-commutative D2 brane discs; this nomenclature is to be regarded somewhat
arbitrary. The energy of these configurations is
\bb\label{v2}
V_{\mbox{disc}}=-\frac{\lk(2 \pi \ls^2 \re)^2}{6 \gs\ls} n\lk(n^2-1\re) M^2\ ,
\ee
which is lower than~\pref{v1} for fixed $n$. 
Issues of classical stability regarding the non-commutative disc and sphere
will be addressed in the next section.

\section{The spectrum of perturbation modes}
\label{modes}

We first focus on the non-commutative sphere configuration. We perturb 
the solution by a general traceless Hermitian matrix $\e^n$
\bbb\label{perturb}
\Phi^i=\frac{\sqrt{-M}}{2} \tau^i+\e^i\ ,\nonumber \\
\Phi^a=\e^a\ .
\eee
And we study the spectrum of the $N^2-1$
eigenvalues of the equations of motion linearized in $\e$.
Some of the details of
this analysis are presented in Section~\ref{details}. In this section, 
we discuss the results.

The equations for $\e^a$ decouple from those for
$\e^i$. Requiring that the $\e^a$ modes are oscillatory leads
to the statement
\bb\label{bound}
M' \ge M\ .
\ee
Note that $M<0$. Negative values for $M$ or $M'$ correspond to
tidal forces pulling the D0
branes away from the origin in the relevant directions of space. 
When the bound~\pref{bound} is satisfied, there are no instabilities in the
six directions transverse to the D2 brane; tachyonic modes would appear only when
$M'$ becomes negative {\em and} is of the same order as $M$. 
To sustain a D2 brane configuration expanding
in the three dimensional subspace, it was found in~\cite{VVSD} that
the negative tidal forces in the three different directions that
we labeled by $i,j$
needed to be of comparative magnitudes.
We then see in~\pref{bound} the
corresponding statement with respect to the other six space dimensions.
When the inequality~\pref{bound} fails, we may then expect
polarization of the D0 branes into D8 branes instead.
If we were to consider the case with anisotropic six dimensional subspace,
we would expect to realize non-commutative D4 and D6 branes.

The dynamics of the $\e^i$ modes is
considerably richer. We have analyzed the spectrum of eigenvalues in
detail first numerically, then identified the analytical pattern underlying it;
we present the results with the help of a series of figures. 

\begin{figure}
\epsfysize=5cm \centerline{\leavevmode \epsfbox{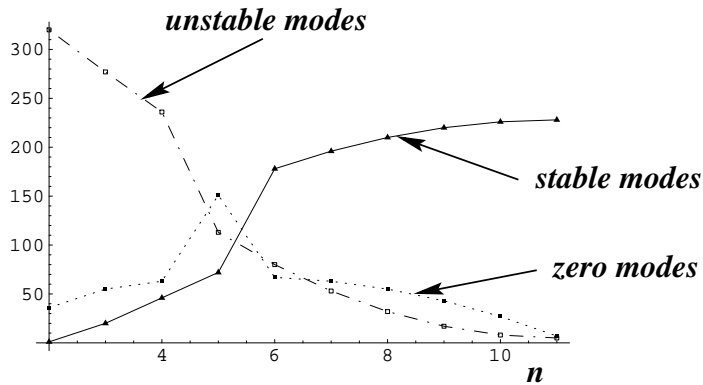}}
\caption{\sl Spectrum of eigenvalues for the non-commutative sphere:
we plot $n$ on the horizontal axis, and count number of
eigenvalues on the vertical.
$N$ is fixed to 11, while $n$, the representation size of the SU(2)
embedded in SU(N), is being varied from 2 to 11. We have also
subtracted from the number of zero modes the dimension
of the space of gauge transformations $n^2-1$.
}
\label{fig1}
\end{figure}

Figure~\ref{fig1} demonstrates how the number of unstable modes decreases
as we increase the size $n$ of the representation, $N$ being fixed.
All modes are physical, as we have modded out the spectrum with the
gauge symmetry of the theory.

\begin{figure}
\epsfysize=5cm \centerline{\leavevmode \epsfbox{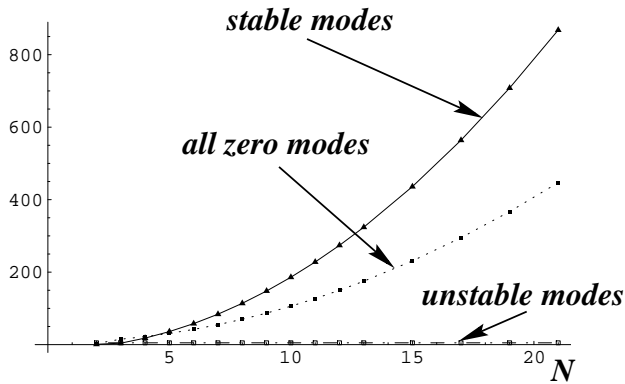}}
\caption{\sl Eigenvalue spectrum of the maximal non-commutative sphere solution: 
$n=N$ is being varied on the horizontal axis, while the number of eigenvalues
is plotted on the vertical. The number of zero modes include gauge transformations;
the number of physical zero modes is equal to seven for all $N$.
}
\label{fig2}
\end{figure}

In Figure~\ref{fig2}, we focus on the maximal representation $n=N$ and vary
$N$. The zero modes include gauge transformations; the number
of physical zero modes is always seven. A striking feature is
that the number of tachyonic modes is also constant; unstable modes
are five in number for all $N$. The rest of the perturbations 
correspond to stable modes. In particular, $3 N-5$ of the Cartan directions
are lifted.
We will appreciate more this phenomenon
when we look at the disc solutions along the same ideas.

It is well-known that the space of $N\times N$ traceless Hermitian
matrices can be organized into the spherical harmonic modes of 
SU(2)~\cite{FUNNEL,WATISPHERE}. 
For a given $N$,
one needs angular momentum modes $l=1,2,\ldots,N-1$; with
$\sum_{l=1}^{N-1} (2 l+1)=N^2-1$.
We will use this basis for 
the matrices $\e^n$ and label the resulting eigenvalue spectrum 
by the harmonic modes they are associated with.

We find that the five unstable modes of the non-commutative sphere solution have
mode number one. They are longest wavelength perturbations and
lie in an SU(2) subalgebra of SU(N), with eigenvectors given by
\bb\label{evec}
\begin{array}{lc|c|c|c|c}
\e^1 \rightarrow & \tau^1  & 0       & \tau^2 & 0      & \tau^3 \\
\e^2 \rightarrow & -\tau^2 & \tau^2  & \tau^1 & \tau^3 & 0      \\
\e^3 \rightarrow & 0       & -\tau^3 & 0      & \tau^2 & \tau^1 
\end{array}
\ee
The first two columns correspond to squashing the sphere into a pancake, one in
each of three directions; in this respect,
only two are linearly independent.
The last three modes are ``rotations with the wrong sign''\footnote{The three
true rotation modes are zero modes, and consist of gauge transformations.}.
The eigenvalues for all these unstable modes are given by
$\omega^2=2 M$. The breathing 
mode, $\e^i\rightarrow\tau^i$, which also appears at the longest wavelength, is stable,
with frequency $\omega^2=- 4 M$.

All of the seven physical zero modes are quadrapole moments, with harmonic mode two.
Any short wavelength perturbation, shorter than harmonic mode one, is otherwise a
stable mode. For example, plucking several D0 branes does not compromise
the global coherence of the configuration. All instabilities arise at the longest
possible wavelengths, at harmonic mode one. The spectrum of the stable modes
is described in Section~\ref{details}.

\begin{figure}
\epsfysize=5cm \centerline{\leavevmode \epsfbox{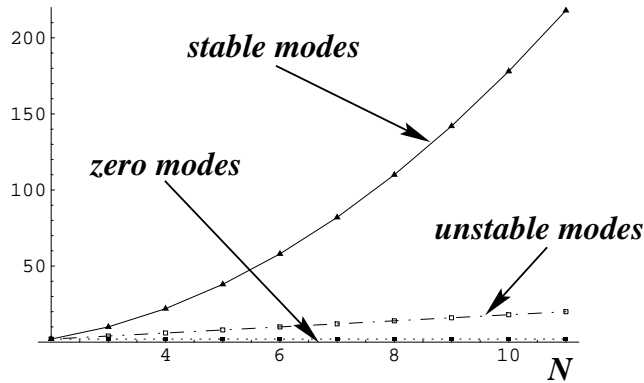}}
\caption{\sl The non-commutative disc solution: we consider maximal embedding
$n=N$; $N$ is being varied on the horizontal axis. Zero modes include gauge
transformations. The number of physical zero modes is fixed to two.
}
\label{fig3}
\end{figure}

Two interesting features of the tachyonic modes
for the maximal embedding of SU(2) into SU(N)
are worth emphasizing:
first, that their number is independent of $N$;
it is always equal to five; and second, that their masses 
are independent of $N$. One immediate consequence is that the fraction
of unstable channels available for decay, out of the $6(N^2-1)$ dimensional 
phase space available for fluctuations, is parametrically small with large $N$.
To underscore the peculiarity of this pattern,
let us contrast the situation
with the other static solutions to~\pref{eom}. First,
the $n$ dimensional representations, with $n<N$, have a growing
number of unstable modes as a function of increasing $N$; this
number grows as $N^2$. The masses of these tachyonic
modes are all degenerate, scaling as $\omega^2=2 M$. Furthermore,
perturbing the disc solution given by~\pref{disc} in a similar manner, 
we find the spectrum depicted in Figure~\ref{fig3}.
The representation here is maximal, with $n=N$, and $N$ is being varied. We see 
a rise in the number of unstable modes as $2 N-2$ even for the maximal 
embedding. Note that these solutions
have lower energy than the non-commutative
sphere for the same $N$; however, for large $N$,
we see that 
they are characterized with a much larger number of decay channels than the 
maximal sphere.
Furthermore, the tachyonic modes are distributed along the spectrum 
$\omega^2=4 M, 6 M,\cdots, 2 N  M$.

We conclude that all of the static solutions 
presented in the previous Section are classically unstable.
All have tachyonic modes, with varying magnitudes and cardinality.
The timescale for each decay channel is set by $L$ 
(with $M\sim 1/L^2$). As we described above, this is a short time
for the objects to be interesting on timescales set by the evolution
of the center of mass; the scale we denoted by $T$.
Our interest lies in tracing the system
as it transports itself across many curvature scales in spacetime. 
However,
as we will see in Section~\ref{time}, 
a careful analysis leads to the conclusion that,
for large $N$ and small string coupling $\gs$, the non-commutative sphere
solution gets singled out and may be expected to be long-lived.

\section{The shape of the potential}
\label{shape}

\begin{figure}
\epsfysize=12cm \centerline{\leavevmode \epsfbox{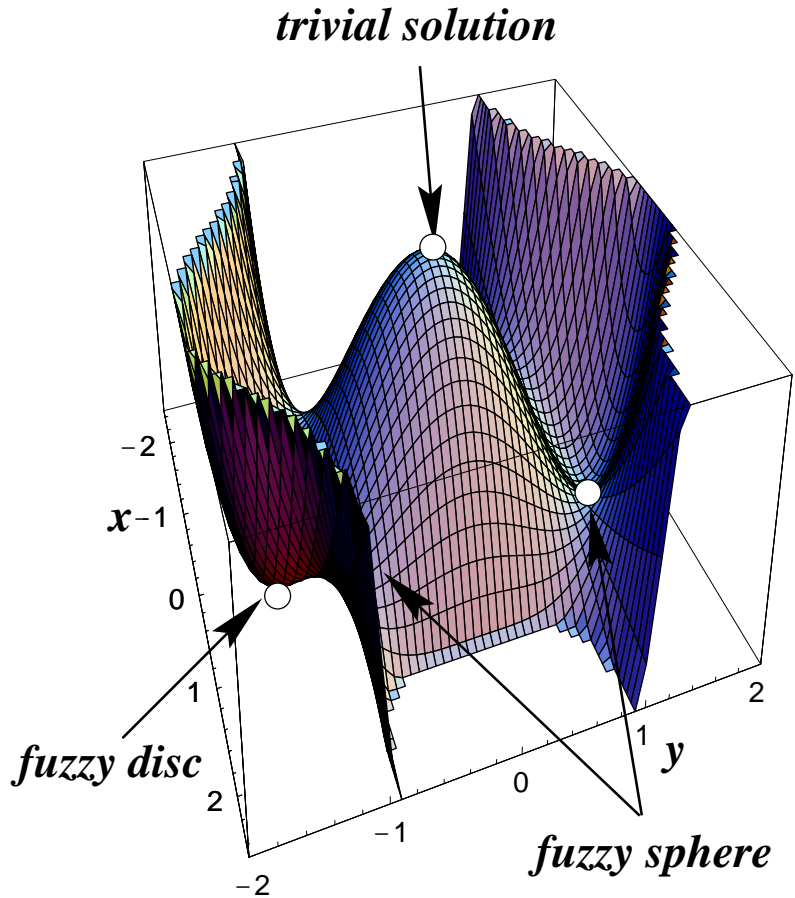}}
\caption{\sl 
A cross-section of the potential energy in~\pref{energy}: the z-axis is
energy, and the cross-section is obtained by 
$\Phi^1=x\ \Phi^1_{(0)}$,
$\Phi^2=y\ \Phi^2_{(0)}$,
$\Phi^3=-y\ \Phi^3_{(0)}$. We have defined $\Phi^i_{(0)}\equiv \sqrt{-M} \tau^i/2$.
}
\label{fig5}
\end{figure}

From our discussion of the spectrum of perturbation modes in the previous
section, it is apparent that our static configurations lie at saddle
points of the potential. 
For the maximal non-commutative sphere,
the phase space of the perturbations about the solution consists of
five unstable tachyonic directions, seven zero modes, and $2 N^2-14$ stable
channels. In an effort to visualize the landscape of this interesting potential
away from its extrema, 
we plot a couple two dimensional cross-sections.
Figure~\ref{fig5} shows three classes of extrema of the potential.
In the center, on a bump, is the trivial solution where the D0 branes sit
on top of each other, \ie\ $\Phi^n=0$. 
The saddle points correspond to the maximal non-commutative sphere
solutions. While the local minima appearing at the bottom of paraboloids
are the $n=N$ disc solutions. It is important to emphasize that this is a two
dimensional cross section; and the many directions of instability 
associated with the disc solutions are suppressed. The infinite well appearing
in the figure leads to a Cartan direction in SU(N); hence to evaporation 
into widely separated D0 branes.

\begin{figure}
\epsfysize=8cm \centerline{\leavevmode \epsfbox{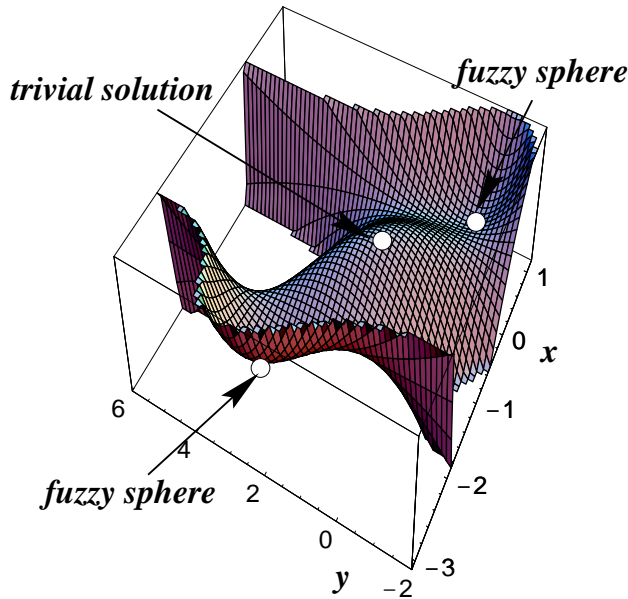}}
\caption{\sl
Another cross-section of the potential energy: this viewpoint
is obtained by 
$\Phi^1=\Phi^1_{(0)}+x\ \Phi^1_{(0)}$,
$\Phi^2=y\ \Phi^2_{(0)}+x\ \Phi^2_{(0)}$,
$\Phi^3=\Phi^3_{(0)}+x\ \Phi^3_{(0)}$; with
$\Phi^i_{(0)}\equiv \sqrt{-M} \tau^i/2$.
}
\label{fig6}
\end{figure}

Another cross-section of this space is shown in 
Figure~\ref{fig6}. This depicts a more ``generic'' viewpoint for large $N$.
The hill in the middle is the trivial solution,
while the ditches on its sides correspond to maximal non-commutative sphere
solutions. The five unstable directions about the sphere solution do not appear
in this cross-section.

These graphical depictions of the energy tell us that the system 
is a complex one; with many extrema of the potential energy
scattered in interesting patterns,
in addition to the gorges that lead to evaporation into
D0 branes. In the next section, we address questions pertaining
to the dynamics of the system in this landscape.

\section{The timescale of decay}
\label{time}

Given that the extrema of the potential energy
corresponding to the polarized states
are saddle points, the obvious conclusion is that
these states, being classically unstable, 
will decay; in particular, they would tend to evaporate to widely separated D0
branes. The distribution of
these extrema in the landscape of the potential reflects
nevertheless certain interesting characteristics of the underlying
dynamics; it is information about the balance
of forces amongst the D0 branes and background supergravity fields. Furthermore,
within the context of the classical instability of these solutions, 
there remains to determine the role of these states in the dynamics
of the system. In particular, 
we would like to find the likelihood that such
polarized states may nucleate; and if they do, to estimate the typical
timescales they live, before evaporating into D0 branes.
We will see below that there is indeed regimes 
where these extrema of the potential play a part in the history
of the evolution of the D0 branes, as the branes cascade their way down the
potential energy.

For the sake of concreteness, we first concentrate on the following scenario.
We consider a background field configuration such that, in some region of space,
spacetime curvatures are small with respect to the string scale.
We project $N$
D0 branes with some initial velocity from this region
into a direction where gradients of
background fields become larger. We also arrange our D0 branes into a polarized
D2 brane configuration corresponding to the $n=N$ solution of~\pref{fuzzy},
where $M$ relates to the initial local curvature scale.
The center
of mass of the system will follow a geodesic in a rescaled metric~\cite{VVSD};
while, in the freely falling frame, 
$M$ evolves on timescales of order $T$.
At every moment in time, we denote the
typical length scale over which background fields vary as 
$L\sim 1/\sqrt{|M|}$.  The scenario is arranged to be 
adiabatic, with $T\gg L$, at least for the initial 
part of the trajectory of the center of mass.
The eventual outcome is
total evaporation into D0 branes; a roll down the bottomless potential.
On their way on this ill-fated journey, 
the D0 branes may however 
spend a significant amount of time near the initial saddle point or other extrema
of the potential energy; that is,
wherever timescales of dynamics becomes of order $T$.
Extrema characterized by times
much shorter than this scale, and in particular ones of order $L$, 
will be mostly irrelevant to the overall dynamics of the system.
The issue then becomes whether we may encounter basins of ``metastability'' during
this evolution.

Let us try to estimate the time the system
will spend in our initial unstable configuration, using the data about
perturbation modes we collected in Section~\ref{modes} . Classically, considering
the idealized situation whereby the configuration sits initially right on top of the
saddle point, with zero initial momentum (in the center of mass frame!), 
we would immediately conclude that
the system stays there forever. However, quantum mechanically, there will
be fluctuations in the matrix elements of the
coordinates $\Phi^m$ and their canonical momenta. 
These new initial conditions will drive the system away from the saddle point,
down unstable directions. The evolution of the configuration can be treated
classically, except for the input of initial conditions given by the appropriate
quantum fluctuations about the saddle point. We describe the state of the
system at some time $t$ after we let go of it by $\Phi^n(t)$
\bb
\Phi^n(t)=\Phi^n_{(0)}+\varepsilon^n(t)\ ,
\ee
where $\Phi^n_{(0)}$ is the saddle point solution given by~\pref{fuzzy}
with $n=N$, and
$\varepsilon^n(t)$ is the small perturbation evolving the system away from
the extremum. To linearized approximation, the dynamics of the $\varepsilon^n$'s
is that of a collection of decoupled harmonic oscillators, with both real
and imaginary frequencies; the latter being the unstable modes.
The frequencies of all these oscillators scale as $1/L$. Hence, the timescale
of the evolution of the system appears to be short. There are several features
that may however drive this timescale in the other direction. 
One is the fact that the number of unstable directions is independent of $N$;
another is that the mass of the tachyons is also independent of $N$; and
finally, the system is endowed with another dimensionless parameter that sets the scale
for the inertia of the configuration. We will argue below that these three
ingredients conspire to set a regime where the system
can spend significant amounts of time, with respect to scale $T$, near certain
saddle points, such as the one corresponding to our initial configuration.

We will need to devise a criterion for determining whether the system
has moved away from the extremum enough so as to consider it 
different for the initial object, the non-commutative sphere;
regarding it instead a remnant of its decay.
A natural approach is 
to look for fractional changes in gauge invariant observables.
For example, we may look at
\bb\label{feq}
f^2\equiv
\frac{\delta\lk[ \Tr \lk( \Phi^n \Phi^n\re)\re]}
{\Tr \lk( \Phi^n_{(0)} \Phi^n_{(0)}\re)}\ .
\ee
Physically, this is the fractional change in the radius of the 
non-commutative sphere.
We may consider higher moments in $\Phi^n$ as well; but it will become apparent 
in the upcoming discussion
that they will provide us no stringer probes of the situation.

At this point, the details of the unstable modes will become important.
We remind the reader that we have five such modes, all at spherical harmonic
mode number one; \ie\ they are longest wavelength perturbations. The eigenvectors
were given in equation~\pref{evec}. All other modes are stable oscillators, or
one of seven zero modes. The frequencies of stable oscillations increase
parametrically with $N$, as described in Section~\ref{details}. For large
times $t\gg L$, the relevant dynamics is only the exponential evolution of the
unstable modes. Stable and zero modes play a subleading role in estimating~\pref{feq}.

Using~\pref{evec}, we evaluate the fraction~\pref{feq},
keeping only contributions from unstable modes. And
we would like this fraction $f$ to be of order a few percent; this is our choice for a
criterion of decay. We note that an important ingredient in computing~\pref{feq} is 
the statement that different harmonic modes decouple from each other
(see identity~\pref{identity}). 
We consequently find that order $\varepsilon$ contributions
to~\pref{feq} vanish identically because of the particular form of the unstable
eigenvectors and subsequent cancelations\footnote{The only contribution to~\pref{feq}
to order $\varepsilon$ comes from the breathing mode $\varepsilon^i\sim \tau^i$, 
which is a stable oscillator
with frequency $\sim 1/L$.}. To quadratic order, we easily obtain the leading
contribution for large times $t\gg L$
\bb
\delta \Tr \lk( \Phi^n \Phi^n\re)\rightarrow \frac{8}{3} N \lk(N^2-1\re) 
\varepsilon_u^2(t)\ .
\ee
We have assumed comparable initial conditions for all five unstable modes,
and we have represented the evolution of these modes with a single function
$\varepsilon_u^2(t)$. The latter is given by
\bb
\varepsilon_u(t)=\Delta x\ \ch (t/L)+L\ \gs\ \ls\ \Delta p\ \sh(t/L)\ .
\ee
$\Delta x$ and $\Delta p$ are respectively the fluctuations in position
and momentum of a tachyonic mode at the saddle point, used as initial
conditions at $t=0$ for the classical evolution of the mode. Furthermore,
we expect that the initial wavepacket, corresponding to the system
siting at the saddle point with minimal initial momentum, saturates 
the uncertainty bound $\Delta p\sim 1/ \Delta x$. Putting things together,
we find that for large times $t\gg L$,
\bb\label{f1}
f\sim \frac{L \Delta x}{\ls^2}\lk(1+\frac{\gs \ls L}{(\Delta x)^2}\re) e^{t/L}\ .
\ee
Note that, despite the fact that the dimension of phase space available for
quantum fluctuations is of order $N^2$, there is no $N$ dependence in~\pref{f1}.
This is a direct result of the form of the spectrum of eigenvalues
discussed in Section~\ref{modes}.

From the work of~\cite{DKPS}, we know that the expected spread $\Delta x$ of the
wavepacket of each mode is given by the eleven dimensional
Planck scale
\bb
\Delta x \sim \lp^{(11)}\sim \gs^{1/3} \ls\ .
\ee
We then obtain
\bb
f\sim \frac{L \gs^{1/3}}{\ls}\lk(1+\frac{L \gs^{1/3}}{\ls}\re) e^{t/L}\ .
\ee
Taking the regime 
\bb
\frac{L \gs^{1/3}}{\ls}\ll 1\ ,
\ee
we estimate the time $\tau_0$ the non-commutative sphere spends near this saddle
point of the potential energy\footnote{The term $\ln f$ is of order one
and can be dropped.}
\bb\label{lifetime}
\tau_0\sim L \ln \lk( \frac{\ls}{L \gs^{1/3}}\re)\ .
\ee
Hence, it is possible for $\tau_0\sim T$, with $T\gg L$, if $\gs$ is taken small
enough. 
In such a scenario, as $L$ evolves on timescales of order $T$,
the D0 branes would continue to track the initial saddle point.
The polarized configuration may then be considered long-lived, disintegrating 
slowly into incoherent D0 branes.

Repeating our analysis for the case of the disc and 
non-commutative sphere with $n<N$, and using
our results for the corresponding eigenvalue spectra given in
Section~\ref{details}, it is easy
to see that we get 
$\tau \sim (L/\sqrt{N}) \ln (\ls/(L N^{1/2} \gs^{1/3}))$ 
and 
$\tau \sim L \ln (\ls/(L N \gs^{1/3}))$ respectively.
This arises from the dependence of the number of
unstable modes, and the corresponding tachyonic masses, on $N$; for larger $N$, there
are more channels of decay, with larger masses for the tachyons in units of $L$.
Without going into the details, it is sufficient to observe that the disc and
$n<N$ spherical solutions will then have parametrically 
shorter ``lifetimes'' than~\pref{lifetime} for large $N$.

This distinguishing property of the $n=N$ non-commutative sphere solution
becomes more important when we want to determine the probability
the system will find a ``basin'' of metastability if it were to start
its journey with more generic initial conditions. 
Noting that the dimension of
the Cartan subalgebra scales as $N$, while the dimension of the
non-abelian channels scales as $N^2$, and that the maximal sphere solution 
is at a minimum in almost all of those directions (including the Cartan ones),
we expect that, for larger values of $N$,
the system would be more likely to find the non-commutative
sphere configuration in the phase space available to it.
Furthermore, the $n<N$ and disc solutions become shorter lived in this limit. 
While the dynamics can be get very complicated,
a large $N$ limit leads to a simplification in this sense.

Focusing on the non-commutative sphere solution with maximal embedding $n=N$
as the prime candidate for a long-lived polarized configuration,
we note however that the scaling in~\pref{lifetime} puts 
the string coupling at a disadvantage in the competition between the various
parameters; in particular, a logarithm of $\gs$ is to compensate a power law in $L/T$.
Hence, the regime we may envisage where the 
maximal non-commutative sphere solution 
will play a salient role in the evolution of the system involves a delicate 
hierarchy between different scales;  which we need 
to discuss carefully both from the point of view
of the embedding string theory, as well as that of
the quantum mechanics
that describes the dynamics of the D0 branes.

\begin{figure}
\epsfxsize=10cm \centerline{\leavevmode \epsfbox{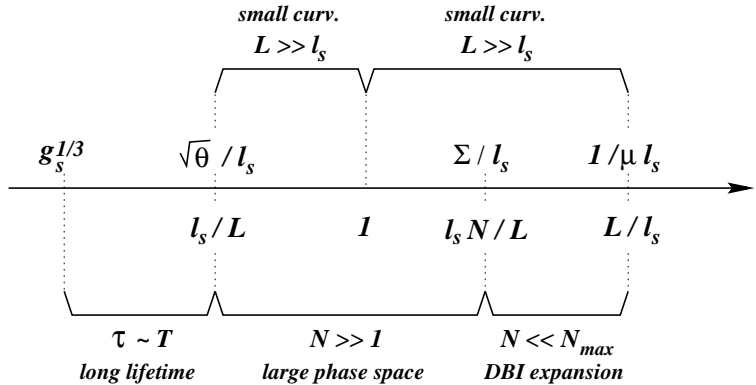}}
\caption{ \sl The hierarchy between the various scales of the problem.
The axis is labeled by various length scales in string units. $\sqrt{\theta}$ denotes
the ``UV cutoff'' in the non-commutative worldvolume theory of the
D2 brane sphere; $\Sigma$ is the IR cutoff due to the topology of
the configuration; and $\mu$ is the energy scale at which the effective
0+1 dimensional Yang-Mills coupling is evaluated at in the expression for
the lifetime of the non-commutative sphere~\pref{lifetimetwo}. 
All brackets indicate a large hierarchy
between the corresponding scales; the associated labels indicate the reasons
for the needed hierarchies. 
}
\label{fig7}
\end{figure}

Figure~\ref{fig7} depicts all scales in the problem written with respect to IIA
string theory and D0 brane quantum mechanics variables. For
large N, the relevant coupling in the 0+1 dimensional
Yang-Mills theory is
\bb
(\geff(\mu))^2\equiv\frac{\gym^2 N}{\mu^3}\equiv \frac{\gs\ N}{(\ls\mu)^3}\ ,
\ee
where $\mu$ is the energy scale at which the coupling is measured.
We then can write
\bb\label{lifetimetwo}
\tau_0\sim L\ \ln\lk(\frac{N}{(\geff(1/L))^2}\re)\ .
\ee
For a perturbative regime in this 
effective coupling, the lifetime $\tau_0$ is then pushed toward larger values.
Hence, from the quantum mechanics point of view, we need a weakly coupled
system, in the large $N$ limit, so that $\tau\sim T\gg L$. 
Note however that we have an upper bound on $N$
coming from the requirement that the expansion of the DBI action is self-consistent;
this bound is given by~\pref{Nbound}, or $N\ll N_{max}\sim L^2/\ls^2$.
Of course, we also need weak background curvature scales, $L/\ls\gg 1$.
In type IIA background geometry language, 
the regime of interest is somewhat similar
to the decoupling limit of the 0+1 Yang-Mills theory; \ie\ we want
small coupling $\gs\rightarrow 0$, large $N$, and
the coupling constant of the quantum mechanics
held fixed (and small). The summary of all these scales and the relevant bounds is
shown in the figure.

\section{Reflections}
\label{reflections}

It is instructive to push our analysis to the limits of its validity.
Looking at~\pref{lifetime}, we can see that 
the polarization effect becomes more important when the center of mass is
``moving fast'', $L\sim T$. The non-commutative state is also longer
lived in regions of space with higher curvatures, near stringy 
length scales $L\sim \ls$. Even though our description clearly breaks
down before these bounds are reached, we may learn from these
limits that polarization effects by gravitational tidal
forces may be most relevant, for example, near singularities.
We may also expect that the coherence of the 
polarized configuration will get compromised
as the center of mass slows down or when entering into regions of spacetime with
progressively smaller curvature scales.

In summary, we have argued for a regime where we expect 
the D0 branes to cascade
down the potential energy landscape, with possibly short visits to the various
extrema with $n<N$, and settling into the $n=N$ non-commutative sphere
configuration for significant amounts of time; finally,
the system will find the Cartan directions of commuting matrices, 
and we are left with a collection of widely separated D0 branes
without any collective coherence; which we may regard as the by-product
of evaporation of the maximal non-commutative sphere. It is important
to emphasize the role of the large $N$ limit in this picture. The regime
of small string coupling increases the inertia of all configurations 
corresponding to extrema of the potential energy. In this respect, the
statement regarding a long lifetime for these classically unstable states is trivial,
being relevant for all the polarized configurations. The large $N$ limit however
singles out the maximal non-commutative sphere solution by rendering the
other extrema shorter lived; while making the dimension of the phase space
in the non-abelian directions parametrically big with respect to that of the 
Cartan channels. This statement makes use of the details of the physics
associated with the maximal non-commutative sphere solution; in particular,
the facts that the number of unstable channels is independent of $N$, that
the corresponding masses of the tachyons are also $N$ independent, and that the
instabilities arise at the longest wavelengths only.

Furthermore, it is interesting that the corresponding spherical D2 brane
configuration is not unstable with respect to short wavelength 
perturbations; in contrast to the usual classical
instabilities associated with bosonic membranes (see, for example,~\cite{MATREV}).
The phenomena we are studying may be generic with respect
to couplings of non-commutative configurations to curved spaces. The 
$M_{mn}\Phi^n \Phi^m$ term is a typical representative of many such couplings
involving symmetrized products of the D0 brane matrix coordinates. This
class of interactions arises in the DBI action due to the prescription 
introduced in~\cite{MEYERS,GARMEY1,GARMEY2} for replacing the
spacetime dependence of background fields with the matrix coordinates of the
D branes.

Given that the regime of interest is somewhat similar to the Matrix theory
decoupling regime~\cite{MAT1}, and noting some of
the similarities between  our polarized configuration 
and Matrix black holes~\cite{HORMART,BFKSHole}, it would be interesting to understand
the relevance of all this to Matrix theory in curved spaces. In particular, extrema
of actions describing an object
as a probe in classical background fields necessarily contain qualitative
information about the reverse effect, involving the back-reaction of the probes
on the background fields.

Given that the potential energy is bottomless, one would also like to understand
the relevance of
decay through quantum mechanical tunneling. The corresponding
dynamics would be naturally governed by $\hbar$, and its importance
to the overall picture would be determined by the relative sizes between $1/N$
and $\hbar$.

Finally, we may ask about explicit realizations of this effect of
polarization by gravitational tidal forces. One system that is a candidate
is the background geometry of near-extremal D6 branes. It can be shown
that in this scenario we may expect the D0 branes to blow up into 
non-commutative D2 branes as they approach the finite area horizon. 
It may be instructive to understand
the genericity of such an effect by exploring classes of
background geometries.

\section{Some details}
\label{details}

In this section, we collect some of the details involved in analyzing
the spectrum of eigenvalues of perturbations about the maximal non-commutative
sphere solution. A similar analysis can be done for other extrema
of the potential. Using~\pref{perturb} in~\pref{eom}, we obtain easily
\bb\label{ea}
\ddot{\e}^a=- 2 M' \e^a 
+\frac{M}{4} \C{\tau^b}{\C{\tau^b}{\e^a}}\ ,
\ee
\bb\label{ei}
\ddot{\e}^i=-2 M \e^i 
+\frac{M}{4} \lk(
\C{\tau^j}{\C{\tau^j}{\e^i}}
+\C{\tau^j}{\C{\e^j}{\tau^i}}
+\C{\e^j}{\C{\tau^j}{\tau^i}}
\re)\ .
\ee
The perturbation modes can be written in a basis of harmonic modes as
\bb
\e^n=\sum_{l=1}^{N-1} \e_{i_1\cdots i_l} \tau^{i_1\cdots i_l}\ ,
\ee
where~\cite{FUNNEL,WATISPHERE}
\bb
\tau^{i_1\cdots i_l}\equiv\Tless\lk[\Sym\lk[\prod_{j=1}^{j=l} \tau^{i_j}\re]\re]\ .
\ee
By $\Sym$ and $\Tless$ we prescribe to symmetrize the product of Pauli
matrices over $i_1\cdots i_l$, and to project onto the subset traceless on any
pair of indices. For fixed $l$, we have $2l+1$ independent traceless Hermitian
matrices $\tau^{i_1\cdots i_l}$.
We note the following useful identity
\bb\label{identity}
\Tr\lk[\tau^{i_1\cdots i_l} \tau^{j_1\cdots j_m}\re]=0\ \ \ \mbox{for}\ \  m\neq l\ ;
\ee
\ie\ different harmonic modes decouple from each other.
We also find the following commutation relations
\bb
\C{\tau^j}{\tau^{i_1\cdots i_l}}=2\ i\ \sum_a  \varepsilon_{j i_a m}
\tau^{i_1\cdots m \cdots i_l}_{i_a \rightarrow m}\ ,
\ee
\bb
\C{\tau^i}{\C{\tau^j}{\tau^{i_1\cdots i_l}}}=
-4\sum_{a\neq b}\varepsilon_{j i_a m} \varepsilon_{i i_b n}
\tau^{i_1\cdots m \cdots n \cdots i_l}_{i_a\rightarrow m; i_b\rightarrow n}
+4 l \delta_{ij} \tau^{i_1\cdots i_l}
-4 \sum_a \delta_{i_a i} 
\tau^{i_1\cdots j \cdots i_l}_{i_a\rightarrow a}\ .
\ee
The notation is as follows: $\tau^{i_1\cdots m \cdots i_l}_{i_a \rightarrow m}$
means $\tau^{i_1\cdots i_l}$ with the index $i_a$ replaced by $m$.
Substituting these relations in~\pref{ea}, we get
\bb
{\ddot{\e}}^a_{i_1\cdots i_l}=
\lk( M l (l+1)- 2 M'\re){\e}^a_{i_1\cdots i_l}\ .
\ee
We are then immediately led to the classical stability condition
\bb
M'\ge M\ .
\ee
In the three dimensional subspace where the D0 brane may expand, 
we get from~\pref{ei} 
\bb
{\ddot{\e}}^i_{i_1\cdots i_l}=
V^{i;i_1 \cdots i_l}_{j;j_1 \cdots j_l}{\e}^j_{j_1\cdots j_l}\ ,
\ee
with
\bb\label{bigv}
V^{i;i_1 \cdots i_l}_{j;j_1 \cdots j_l}\equiv
M\lk(
(l-2) \delta_{ij} \Delta
+(l-2) \sum_a \delta_{i j_a} \Delta_{j_a\rightarrow j} 
+(l+1) \sum_a \delta_{j j_a} \Delta_{j_a\rightarrow i}
\re)\ .
\ee
We have defined
\bb
\Delta\equiv \delta_{i_1 j_1} \cdots \delta_{i_l j_l}\ ;
\ee
and $\Delta_{j_a\rightarrow j}$ means $\Delta$ with the corresponding substitution
in one of the indices. Expression~\pref{bigv} is to be symmetrized
in the indices $\{i_1\cdots i_l\}$ and projected on the traceless subspace. 

We then have an algorithm for computing the
matrix~\pref{bigv}. Diagonalizing it, we are
led to a spectrum of eigenvalues for the perturbations modes.
For a given $N$, we need to collect together all such eigenvalues
with $l<N$\mbox{ }\footnote{In this harmonic basic, the perturbation
matrix is block diagonal with respect to different harmonic mode sectors.}.

We have implemented this analysis using a computer.
Diagonalizing the resulting matrices numerically, we obtained data
about the spectrum. An analysis of the patterns in this spectrum
inspired us to formulate the analytical description of the solution. 
One immediate and interesting
observation is that all the eigenvalues appear to be integer multiples of $M$
(with an exception to be noted below). Furthermore,
a simple, yet non-trivial, pattern emerges.
Some of the results were summarized in Section~\ref{modes}. Here, we describe
a few more details.

For perturbations about the maximal non-commutative sphere solution, we obtain:
\begin{itemize}
\item Seven physical zero modes at harmonic mode number two.
\item Five tachyonic modes at harmonic mode number one, with mass squared $2 M$.
\item Stable modes appear at all
values of the harmonic numbers. The spectrum is interesting and is depicted
in Figure~\ref{figbars}. These modes can me labeled with a sequence
of integers $z=2,3,\ldots$ such that the frequencies are given by
\bb
\omega_z^2=-M(z^2+z-2)\mbox{   with degeneracy   }d_z=4 z+2\ .
\ee
\begin{figure}
\epsfysize=6cm \centerline{\leavevmode \epsfbox{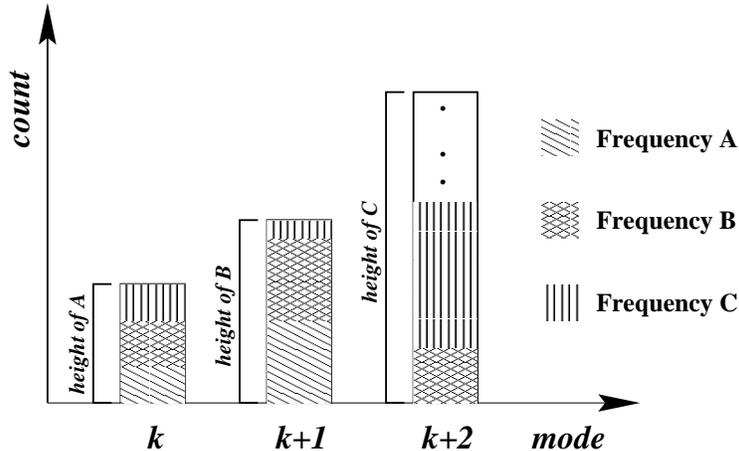}}
\caption{\sl A schematic depiction of 
the distribution of stable perturbations across harmonic modes.
The height of each bar reflects the number of stable eigenvalues in
the corresponding harmonic mode. Each value of a positive frequency 
is represented by some shading pattern. The degeneracy of a given
eigenvalue is proportional to the height of the corresponding shading
pattern. These heights are in one to one
correspondence with the total height of a bar with some fixed harmonic mode.
}
\label{figbars}
\end{figure}
For a fixed value for $z$, the modes are distributed across a complex pattern of
harmonic modes. Remarkably, the number of all stable modes
appearing within a fixed harmonic mode $m$, with $m\ge 3$, is given by $d_m$.
This pattern breaks down
for dipole and quadrapole moments, but appears to be sustained for all
higher ones. It may be an interesting mathematical problem 
to identify the mechanism responsible for this pattern of degeneracies.
\end{itemize}

For the $n<N$ sphere solution, 
the spectrum is complex and, except for a few features,
we cannot describe it analytically. We make the following observations:
\begin{itemize}
\item All frequencies are integer multiples of $M$
only for {\em odd} $N$. Other details 
are also correlated with whether $N$ is divisible by two.
\item The number of unstable modes increases as $N^2$, for fixed $n$.
The masses of these tachyonic modes are the same; we get $\omega^2=2 M$.
\item The number of stable modes increases as $N$.
\end{itemize}

For the maximal $n=N$ disc solution, we find
\begin{itemize}
\item  The number of physical zero modes is always equal to two.
\item The number of unstable modes increases as $2 N-2$.
The masses of these tachyonic modes appear as 
$\omega^2=4 M, 6 M, \cdots, 2 N M$.
\item The number of stable modes increases as $N^2$.
\end{itemize}


\paragraph{\bf Acknowledgments:}
I am particularly grateful to M.~Spradlin and A.~Volovich for
drawing my attention to the problem of instability and for useful discussions.
I also thank P.~Argyres, T.~Becher, M.~Moriconi, H.~Tye, and
E.~Yuzbashyan for many helpful discussions.
This work was supported by NSF grant 9513717.

\providecommand{\href}[2]{#2}\begingroup\raggedright\endgroup


\end{document}